\documentclass[runningheads]{llncs}
\usepackage{graphicx}
\usepackage{comment}
\usepackage{amsmath,amssymb} % define this before the line numbering.
\usepackage{color}

\newcommand{\etal}{\textit{et al}. }
\newcommand{\vs}{\textit{vs}. }

\newcommand{\eg}{\textit{e}.\textit{g}. }
\usepackage{subcaption}
\usepackage{pifont}
\usepackage{multirow}
\usepackage{booktabs}
\usepackage{subeqnarray}
\usepackage{tabularx}
\usepackage{epstopdf}

\begin{document}
% \linenumbers
\pagestyle{headings}
\mainmatter
\def\ECCVSubNumber{95}  % Insert your submission number here

\title{Residual Feature Distillation Network for Lightweight Image Super-Resolution } % Replace with your title

%\begin{comment}
\titlerunning{Residual Feature Distillation Network} 
\authorrunning{Jie Liu, Jie Tang, Gangshan Wu} 
\author{Jie Liu\and
Jie Tang\thanks{Corresponding Author}\and
Gangshan Wu}
\institute{State Key Laboratory for Novel Software Technology, Nanjing University, China\\
{\tt\small jieliu@smail.nju.edu.cn, \{tangjie,gswu\}@nju.edu.cn}
}
%\end{comment}
%******************

\begin{comment}
\titlerunning{Abbreviated paper title}
\author{First Author\inst{1}\orcidID{0000-1111-2222-3333} \and
Second Author\inst{2,3}\orcidID{1111-2222-3333-4444} \and
Third Author\inst{3}\orcidID{2222--3333-4444-5555}}
%
\authorrunning{F. Author et al.}
\institute{Princeton University, Princeton NJ 08544, USA \and
Springer Heidelberg, Tiergartenstr. 17, 69121 Heidelberg, Germany
\email{lncs@springer.com}\\
\url{http://www.springer.com/gp/computer-science/lncs} \and
ABC Institute, Rupert-Karls-University Heidelberg, Heidelberg, Germany\\
\email{\{abc,lncs\}@uni-heidelberg.de}}
\end{comment}
%******************
\maketitle

\begin{abstract}
    Recent advances in single image super-resolution (SISR) explored the power of convolutional neural network (CNN)
    to achieve a better performance. Despite the great success of CNN-based methods, it is not easy to apply these
    methods to edge devices due to the requirement of heavy computation. To solve this problem, various fast and 
    lightweight CNN models have been proposed. The information distillation network is one of the state-of-the-art
    methods, which adopts the channel splitting operation to extract distilled features. However, it is not clear
    enough how this operation helps in the design of efficient SISR models. In this paper, we propose the feature 
    distillation connection (FDC) that is functionally equivalent to the channel splitting operation while being more
    lightweight and flexible. Thanks to FDC, we can rethink the information multi-distillation network (IMDN) and
    propose a lightweight and accurate SISR model called residual feature distillation network (RFDN). RFDN uses 
    multiple feature distillation connections to learn more discriminative feature representations. We also propose a 
    shallow residual block (SRB) as the main building block of RFDN so that the network can benefit most from residual 
    learning while still being lightweight enough. Extensive experimental results show that the proposed RFDN achieves 
    a better trade-off against the state-of-the-art methods in terms of performance and model complexity. Moreover,
    we propose an enhanced RFDN (E-RFDN) and won the first place in the AIM 2020 efficient super-resolution challenge.
    Code will be available at \url{https://github.com/njulj/RFDN}.
 
\end{abstract}

\section{Introduction}
Image super-resolution~(SR) is a classic computer vision task to reconstruct a high-resolution~(HR) image from
its low-resolution~(LR) counterpart. It is an ill-posed procedure since many HR images can be degraded to the 
same LR image. Image SR is a very active research area where many approaches~\cite{DBLP:conf/cvpr/TimofteRG16,DBLP:conf/cvpr/LedigTHCCAATTWS17} have been proposed to generate
the upscaled images. In this paper, we focus on the problem of lightweight image SR which is needed in 
time-sensitive applications such as video streaming.

Recently, various convolutional neural network (CNN) based methods~\cite{,IDN,MemNet,EDSR,DBPN,SRFBN,RCAN} have been proposed and achieved prominent
performance in image SR. As a pioneering work, Dong~\etal~\cite{SRCNN} proposed the super-resolution convolutional neural
network (SRCNN), which is a three-layer network to directly model the mapping from LR to HR. 
Then, Kim~\etal~\cite{VDSR} pushed the depth of SR network to 20 and achieved much better performance than SRCNN, which
indicates that the quality of upscaled images can be improved with deeper networks. The EDSR~\cite{EDSR} network further
proved this by using more than 160 layers. Although deeper networks increase the quality of SR images, they
are not suitable for real-world scenarios. It is important to design fast and lightweight CNN models that have a better
trade-off between SR quality and model complexity.

To reduce the number of parameters, DRCN~\cite{DRCN} and DRRN~\cite{DRRN} adopted a recursive network that decreases the number of 
parameters effectively by parameter sharing. However, it has to increase the depth or the width of the network
to compensate for the loss caused by the recursive module. These models reduce the model size at the expense
of increased number of operations and inference time. In real-world applications, the number of operations
is also an important factor to consider so that the SR model can be performed in real-time. So, it is better
to design dedicated networks that are lightweight and efficient enough for real-world scenarios. 
%\begin{figure}[t]
%    \centering
%  \includegraphics[width=0.8\linewidth]{Figure/urban100_x4.pdf}  
%  \caption{PSNR \vs Parameters on Urban100 $\times 4$ dataset.}
%\end{figure}
%############################################################################################################
\begin{figure*}[t]
    \centering
  \includegraphics[width=0.8\linewidth]{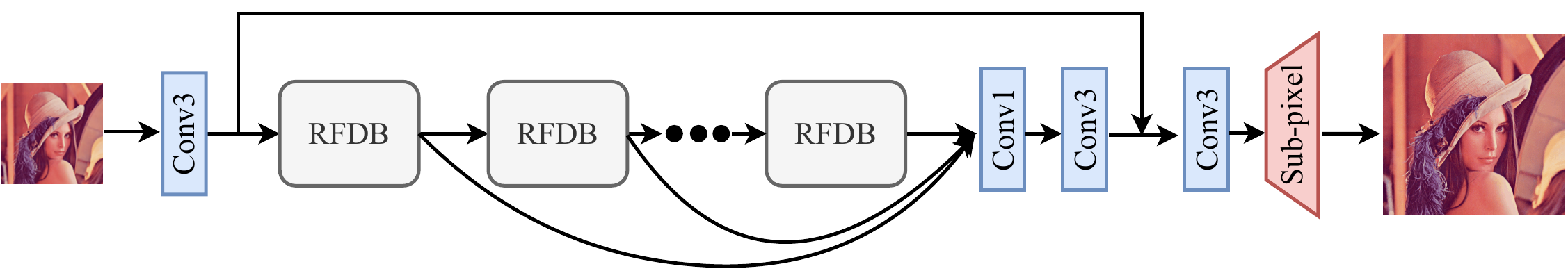}  
  \caption{The architecture of residual feature distillation network (RFDN).}
  \label{fig:rmdn}
\end{figure*}
%############################################################################################################

To this end,
Ahn~\etal~\cite{CARN} proposed the CARN-M for mobile devices by using a cascading network architecture, but it is at the cost
of a large PSNR drop. Hui~\etal~\cite{IDN} proposed an information distillation network (IDN) that explicitly split the 
intermediate features into two parts along the channel dimension, one was retained and the other was further
processed by succeeding convolution layers. By using this channel splitting strategy, IDN can aggregate current
information with partially retained local short-path information and achieve good performance at a modest size.
Later, IMDN~\cite{IMDN} further improved IDN by designing an information multi-distillation block (IMDB) that extracted 
features at a granular level. Specifically, the channel splitting strategy was applied multiple times within
a IMDB. Each time, one part of the features was retained and another was sent to the next step. IMDN has a good
performance in terms of both PSNR and inference time and won the first place in the AIM 2019 constrained image
super-resolution challenge~\cite{AIM2019}. However, the number of parameters of IMDN is more than most of the lightweight SR models
(\eg VDSR~\cite{VDSR}, IDN~\cite{IDN}, MemNet~\cite{MemNet}). There is still room for improvement to be more lightweight.

The key component of both IDN and IMDN is the information distillation mechanism (IDM) that explicitly divides the preceding
extracted features into two parts, one is retained and the other is further refined. 
We argue that the IDM is not
efficient enough and it brings some inflexibility in the network design. It is hard to incorporate identity connections 
with the IDM. In this paper, we will give a more comprehensive analysis of the information distillation mechanism and propose the
feature distillation connection (FDC) that is more lightweight and flexible than the IDM.
%In this paper, we  and propose a functionally identical structure (MSC) but is more flexible than the CSS.  
We use IMDN as the baseline model since it makes a good trade-off between the reconstruction quality and the 
inference speed, which is very suitable for mobile devices. But the IMDN is not lightweight enough and the SR performance
can still be further improved. To build a more powerful fast and lightweight SR model, we rethink the architecture of IMDN and 
propose the residual feature distillation network (RFDN). In comparison with IMDN, our RFDN is much more lightweight by using
the feature distillation connections (FDCs).
%Thanks to this new structure, we reduce a lot of redundant parameters in comparison to IMDN. 
Further more, we propose a shallow residual block (SRB) that uses as the building blocks of RFDN to further improve the SR performance.
The SRB consists of one convoltuional layer, an identical connection and an activation unit at the end. It can benefit from the residual learning~\cite{ResNet}
without introducing extra parameters compared with plain convolutions. It is very easy to incorporate SRB with the feature distillation connection to build a more powerful SR network.

The main contributions of this paper can be summarized as follows:
\begin{enumerate}
    \item We propose a lightweight residual feature distillation network (RFDN) for fast and accurate image super-resolution, which achieves
        state-of-the-art SR performance while using much fewer parameters than the competitors.
    \item We give a more comprehensive analysis of the information distillation mechanism (IDM) and rethink the IMDN network. Based on these new understandings,
        we propose the feature distillation connections (FDC) that are more lightweight and flexible than the IDM.
    \item We propose the shallow residual block (SRB) that incorporates the identity connection with one convolutional block to further improve
        the SR performance without introducing any extra parameters.
\end{enumerate}

\section{RELATED WORK}
Recently, deep learning based models have achieved dramatic improvements in image SR. The pioneering work was done
by Dong~\etal~\cite{SRCNN}, they first exploited a three-layer convolutional neural network SRCNN to jointly optimize the
feature extraction, non-linear mapping and image reconstruction in an end-to-end manner. Then Kim~\etal~\cite{VDSR} proposed
the very deep super-resolution (VDSR) network, which stacked 20 convolutional layers to improve the SR performance.
To reduce the model complexity, Kim~\etal~\cite{DRCN} introduced DRCN that recursively applied the feature extraction layer
for 16 times. DRRN~\cite{DRRN} improved DRCN by combining the recursive and residual network schemes to achieve better
performance with fewer parameters. Lai~\etal~\cite{LapSRN} proposed the laplacian pyramid super-resolution network (LapSRN) to
address the speed and accuracy problem by taking the original LR images as input and progressively reconstructing
the sub-band residuals of HR images. Tai~\etal~\cite{MemNet} presented the persistent memory network (MemNet) for image
restoration task, which tackled the long-term dependency problem in the previous CNN architectures. To reduce the
computational cost and increase the testing speed, Shi~\etal~\cite{ESPCN} designed an efficient sub-pixel convolution to
upscale the resolutions of feature maps at the end of SR mdoels so that most of computation was performed in the 
low-dimensional feature space. For the same purpose, Dong~\etal~\cite{FSRCNN} proposed fast SRCNN (FSRCNN), which employed 
transposed convolution as upsampling layers to accomplish post-upsampling SR. Then Lim~\etal~\cite{EDSR} proposed EDSR
and MDSR, which achieved significant improvements by removing unnecessary modules in conventional residual networks.
Based on EDSR, Zhang~\etal proposed the residual dense network (RDN)~\cite{RDN} by introducing dense connections into the 
residual block. They also proposed the very deep residual attention network (RCAN)~\cite{RCAN} and the residual non-local
attention network (RNAN)~\cite{RNAN}. Dai~\etal~\cite{SAN} exploited the second-order attention mechanism to adaptively rescale features by
considering feature statistics higher than first-order. Guo~\etal~\cite{DRN} developed a dual regression scheme by introducing
an additional constraint such that the mappings can form a closed-loop and LR images can be reconstructed to enhance the performance of SR models.

Despite the great success of CNN-based methods, most of them are not suitable for mobile devices. To solve this 
problem, Ahn~\etal~\cite{CARN} proposed the CARN-M model for mobile scenario through a cascading network architecture.
Hui~\etal~\cite{IDN} proposed the information distillation network (IDN) that explicitly divided the preceding extracted 
features into two parts. Based on IDN, the also proposed the fast and lightweight information multi-distillation
network (IMDN)~\cite{IMDN} that is the winner solution of the AIM 2019 constrained image super-resolution challenge~\cite{AIM2019}.

\section{METHOD}
\subsection{Information multi-distillation block}
%############################################################################################################
\begin{figure}[t]
  \centering
\begin{subfigure}{.24\linewidth}
  \centering
  % include first image
  \includegraphics[width=.7\linewidth, height=0.22\textheight]{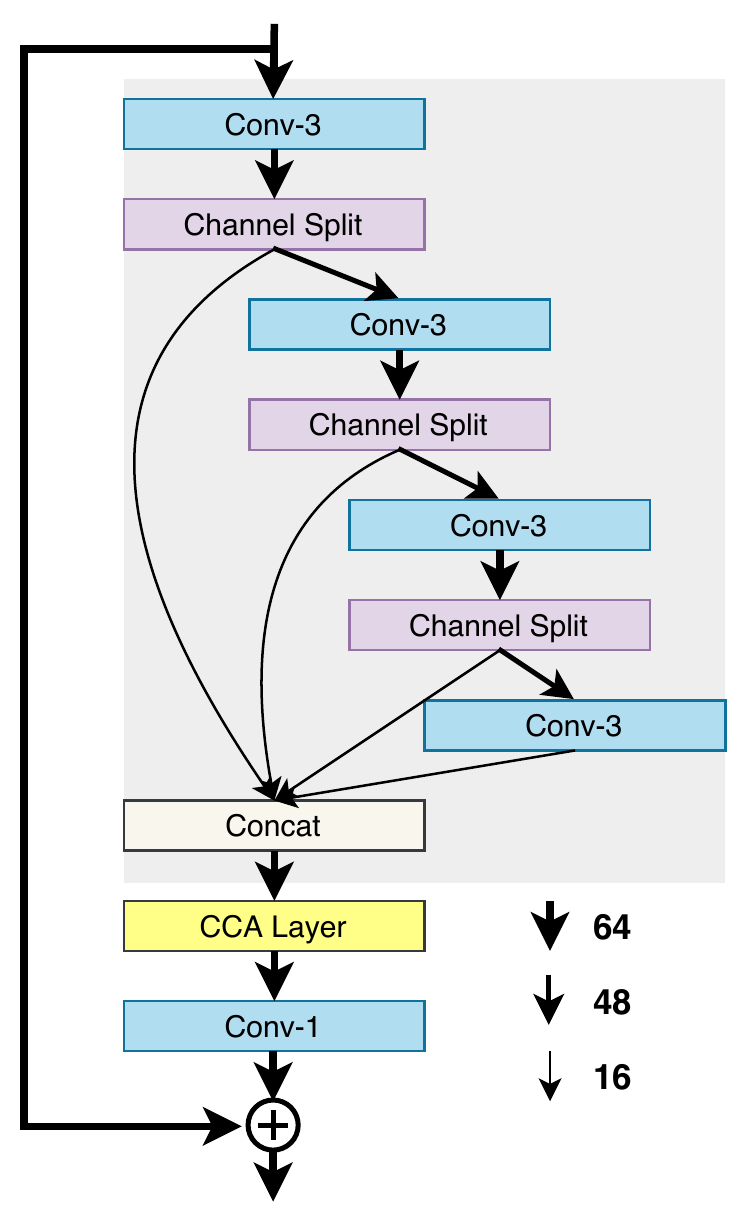}  
  \caption{IMDB}
  \label{fig:imdb}
\end{subfigure}
\begin{subfigure}{.24\linewidth}
  \centering
  % include second image
  \includegraphics[width=.7\linewidth, height=0.22\textheight]{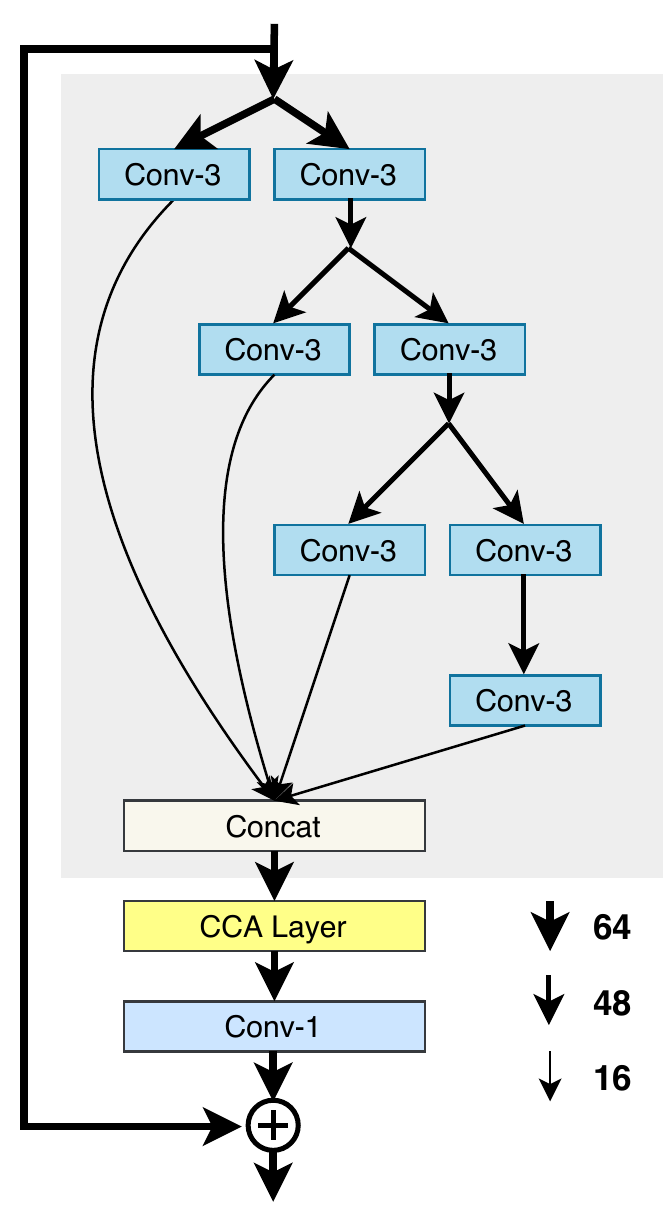}  
  \caption{IMDB-R}
  \label{fig:imdb-r}
\end{subfigure}
\begin{subfigure}{.24\linewidth}
  \centering
  % include first image
  \includegraphics[width=.7\linewidth, height=0.22\textheight]{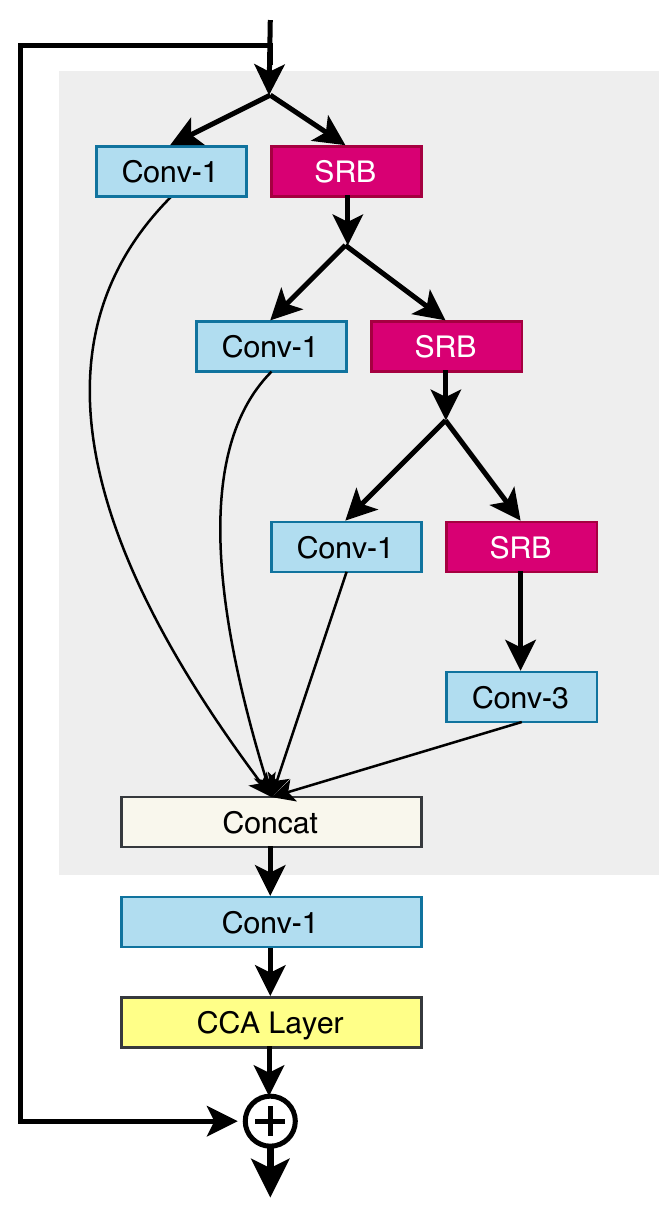}  
  \caption{RFDB}
  \label{fig:rfdb}
\end{subfigure}
\begin{subfigure}{.24\linewidth}
  \centering
  % include second image
  \includegraphics[width=.7\linewidth, height=0.22\textheight]{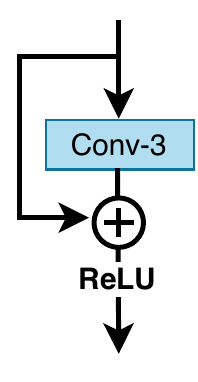}  
  \caption{SRB}
  \label{fig:srb}
\end{subfigure}
\caption{(a) IMDB: the original information multi-distillation block. (b) IMDB-R: rethinking of the IMDB.
(c) RFDB: residual feature distillation block. (d) SRB: shallow residual block.}
\label{fig:imdbs}
\end{figure}

%\begin{figure}[t]
%  \centering
%\caption{Left: the residual feature distillation block (RFDB). Right: the shallow residual block (SRB) included in
%RFDB.}
%\label{fig:rmdb-srb}
%\end{figure}
%############################################################################################################
As shown in Figure~\ref{fig:imdb}, the main part of information distillation block (IMDB)~\cite{IMDN} is a progressive refinement module (PRM), which is marked 
with a gray background.
The PRM first uses a $3\times 3$ convolution layer to extract input features for multiple subsequent distillation
steps. For each step, the channel splitting operation is employed on the preceding features and it divides the input
features into two parts. One part is retained and the other part is fed into the next distillation step.
Given the input features $F_{in}$, this procedure can be described as
\begin{equation} \label{equ:imdb}
\begin{split}
    F_{distilled\_1},F_{coarse\_1} &= Split_1(L_1(F_{in})),\\
    F_{distilled\_2},F_{coarse\_2} &= Split_2(L_2(F_{coarse\_1})),\\
    F_{distilled\_3},F_{coarse\_3} &= Split_3(L_3(F_{coarse\_2})),\\
    F_{distilled\_4} &= L_4(F_{coarse\_3})
\end{split}
\end{equation}
where $L_j$ denotes the $j$-th convolution layer (including the activation unit), $Split_j$ denotes the $j$-th
channel splitting operation, $F_{distilled\_j}$ represents the $j$-th distilled features, and $F_{coarse\_j}$ is
the $j$-th coarse features that will be further processed by succeeding layers. Finally, all the distilled
features are concatenated together as the output of the PRM
\begin{equation}
    F_{distilled} = Concat(F_{distilled\_1}, F_{distilled\_2}, F_{distilled\_3}, F_{distilled\_4})
\end{equation}
where $Concat$ represents the concatenation operation along the channel dimension.

\subsection{Rethinking the IMDB}
Although PRM achieves prominent improvements, it is not efficient enough and introduces some inflexibility
because of the channel splitting operation. The distilled features are generated by $3\times 3$ convolution
filters that has many redundant parameters. Moreover, the feature refinement pipeline 
(along the right branch of the PRM) 
is coupled together with channel splitting operation so that it is hard to use identity connections only for this
pipeline. Next, we will rethink the channel splitting operation and give a new equivalent architecture of the PRM to
tackle the aforementioned problems.

As depicted in Figure~\ref{fig:imdb-r}, the $3\times 3$ convolution followed by a channel splitting layer can be 
decoupled into two $3\times 3$ convolution layers $DL$ and $RL$. The layer $DL$ is responsible for producing the
distilled features and $RL$ is the refinement layer that further processes the proceeding coarse features.
The whole structure can be described as
\begin{equation} \label{equ:imdb-r}
\begin{split}
    F_{distilled\_1},F_{coarse\_1} &= DL_1(F_{in}), RL_1(F_{in})\\
    F_{distilled\_2},F_{coarse\_2} &= DL_2(F_{coarse\_1}),RL_2(F_{coarse\_1}),\\
    F_{distilled\_3},F_{coarse\_3} &= DL_3(F_{coarse\_2}),RL_3(F_{coarse\_2}),\\
    F_{distilled\_4} &= DL_4(F_{coarse\_3})
\end{split}
\end{equation}
Comparing equation~\ref{equ:imdb} with equation~\ref{equ:imdb-r}, we have the following relationships
\begin{equation}
\begin{split}
    DL_1(F_{in}), RL_1(F_{in}) &= Split_1(L_1(F_{in})),\\
    DL_2(F_{coarse\_1}),RL_2(F_{coarse\_1}) &= Split_2(L_2(F_{coarse\_1})),\\
    DL_3(F_{coarse\_2}),RL_3(F_{coarse\_2}) &= Split_3(L_3(F_{coarse\_2})),\\
    DL_4(F_{coarse\_3}) &= L_4(F_{coarse\_3})
\end{split}
\end{equation}
The above equations describe that each group of split operation can be viewed as two convolution layers that work
concurrently. We call this new architecture IMDB-R, which is more flexible than the original IMDB. It has a clearer
view on how the PRM works so that we can get more clues on how to design more efficient SR models.

\subsection{Residual feature distillation block}
Inspired by the rethinking of IMDB, in this section, we introduce the residual feature distillation block (RFDB) that 
is more lightweight and powerful
than the IMDB. In Figure~\ref{fig:imdbs}, we can see that the information distillation operation is actually implemented by 
a $3\times 3$ convolution that compresses feature channels at a fixed ratio. However, we find that it is more
efficient to use the $1\times 1$ convolution for channel reduction as have done in many other CNN models.
As depicted in Figure~\ref{fig:rfdb}, the three convolutions on the left are replaced with $1\times 1$
convolutions, which significantly reduces the amount of parameters. The right-most convolution still uses 
$3\times 3$ kernels. This is because it locates on the main body of the RFDB and it must take the spatial context
into account to better refine the features. For clarity, we call these outer connections feature distillation
connections (FDC).

Despite aforementioned improvements, we also introduce more fine-grained
residual learning into the network. For this purpose, we design a shallow residual block (SRB),
as shown in Figure~\ref{fig:srb}, which consists of a $3\times 3$ convolution, an identity connection
and the activation unit. The SRB can benefit from residual learning without introducing any extra parameters.
The original IMDB only contains mid-level residual connections that are too coarse for the network to benefit most
from the residual connections. In contrast, our SRB enables deeper residual connections and can better utilize
the power of residual learning even with a lightweight shallow SR model. We use the proposed RFDB to build our residual
feature distillation network (RFDN) as will be described in the next section.

\subsection{Framework}
We use the same framework as IMDN~\cite{IMDN}, as shown in Figure~\ref{fig:rmdn}, the residual feature distillation network (RFDN)
consists of four parts: the first feature extraction convolution, multiple stacked residua feature distillation
blocks (RFDBs), the feature fusion part and the last reconstruction block. Specifically, the initial feature
extraction is implemented by a $3\times 3$ convolution to generate coarse features from the input LR image.
Given the input $x$, this procedure can be expressed as
\begin{equation}
    F_0 = h(x)
\end{equation}
where $h$ denotes the coarse feature extraction function and $F_0$ is the extracted features. The next part of
RFDN is multiple RFDBs that are stacked in a chain manner to gradually refine the extracted features.
This process can be formulated as
\begin{equation}
    F_k = H_k(F_{k-1}),k=1,\dots,n
\end{equation}
where $H_k$ denotes the $k$-th RFDB function, $F_{k-1}$ and $F_k$ represent the input feature and output feature
of the $k$-th RFDB, respectively. After gradually refined by the RFDBs, all the intermediate features are 
assembled by a $1\times 1$ convolution layer. Then, a $3\times 3$ convolution layer is used to smooth the 
aggregated features as follows
\begin{equation}
    F_{assemble} = H_{assemble}(Concat(F_1,\dots,F_n))
\end{equation}
where $Concat$ is the concatenation operation along the channel dimension, $H_{asemble}$ denotes the $1\times 1$ 
convolution followed by a $3\times 3$ convolution, and $F_{assemble}$ is the aggregated features. Finally,
the SR images are generated through the reconstruction as follows
\begin{equation}
    y = R(F_{assemble} + F_0)
\end{equation}
where $R$ denotes the reconstruction function and $y$ is the output of the network. The reconstruction process
only consists of a $3\times 3$ convolution and a non-parametric sub-pixel operation.

The loss function of our RFDN can be expressed by
\begin{equation}
    \mathbb{L}(\theta) = \frac{1}{N}\sum_{i=1}^{N}||H_{RFDN}(I_i^{LR})-I_i^{HR}||_1
\end{equation}
where $H_{RFDN}$ represents the function of our proposed network, $\theta$ indicates the learnable parameters of
RFDN and $||.||_1$ is the $l_1$ norm. $I^{LR}$ and $I^{HR}$ are the input LR images and the corresponding
ground-truth HR images, respectively.
\begin{figure*}[t]
    \begin{subfigure}[]{.23\linewidth}
  \centering
  % include first image
  %\includegraphics[width=.8\linewidth]{Figure/base.pdf}  
  \includegraphics[width=.6\linewidth]{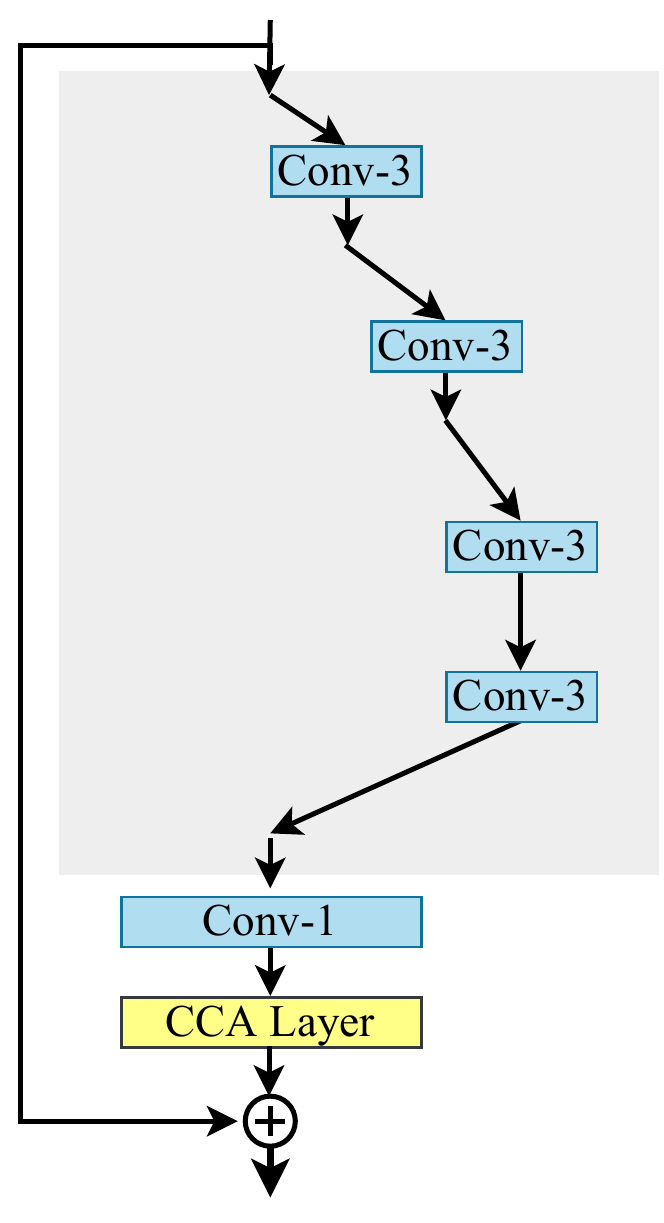}  
  \caption{Base}
  \label{fig:base}
\end{subfigure}
\begin{subfigure}[]{0.23\linewidth}
  \centering
  % include second image
  %\includegraphics[width=.8\linewidth]{Figure/srbs.pdf}  
  \includegraphics[width=.6\linewidth]{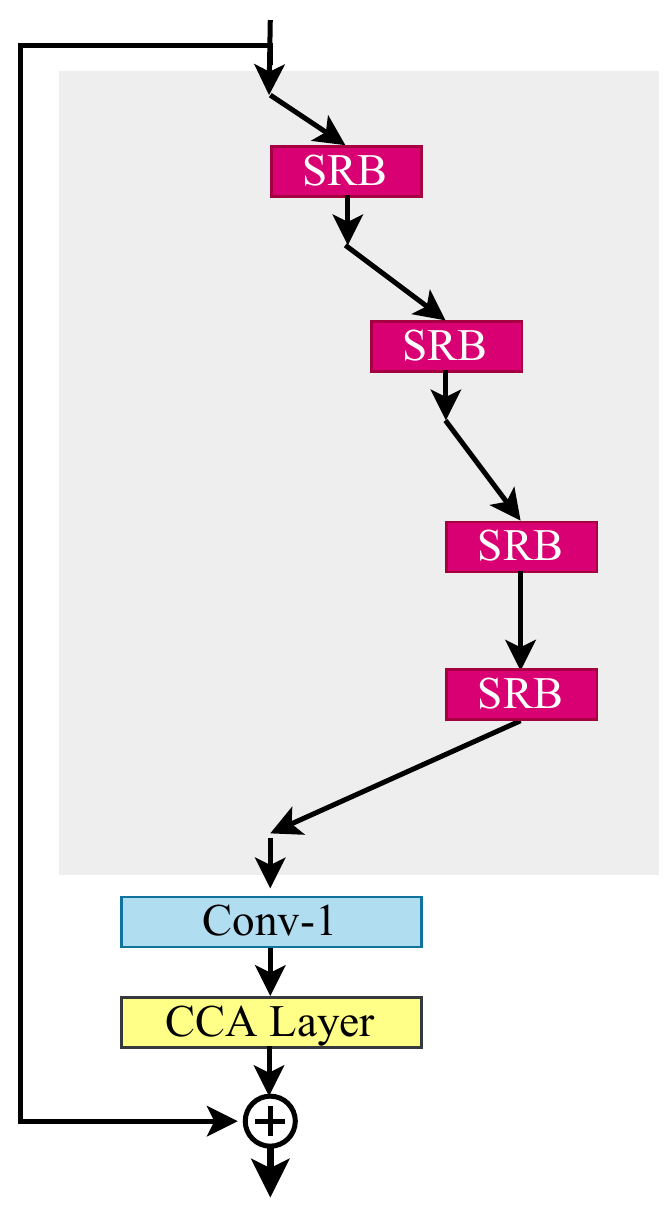}  
  \caption{SRB}
  \label{fig:base_srb}
\end{subfigure}
\begin{subfigure}[]{0.23\linewidth}
  \centering
  % include second image
  \includegraphics[width=.6\linewidth]{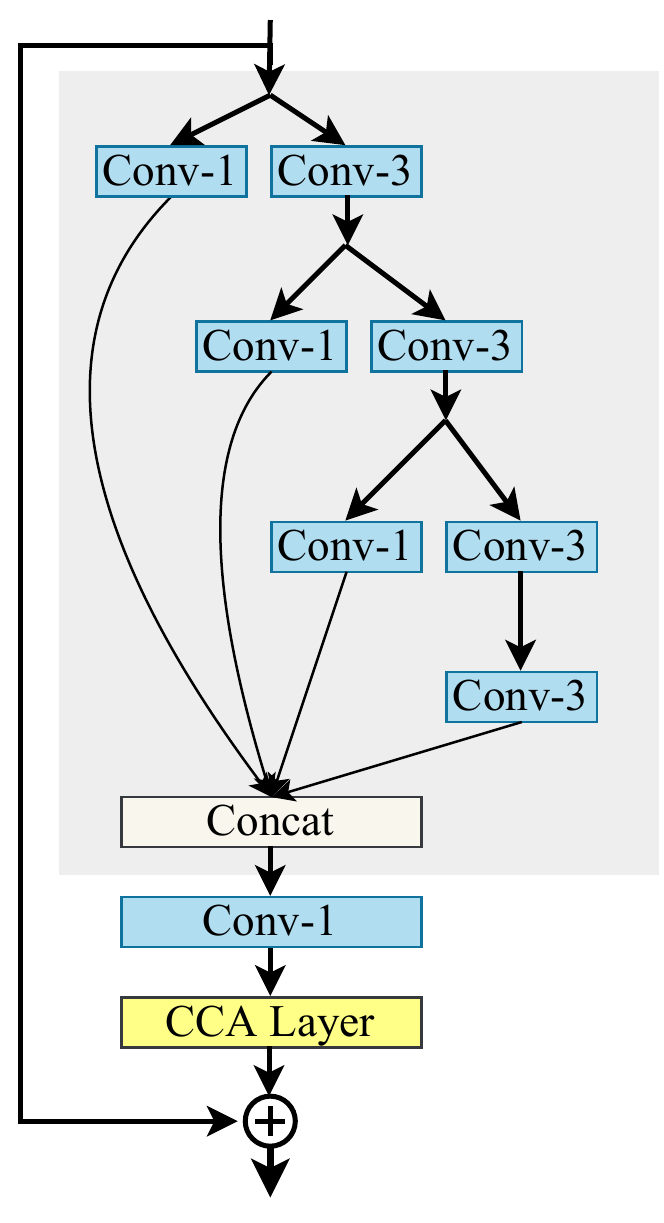}  
  \caption{FDC}
  \label{fig:imdc}
\end{subfigure}
\begin{subfigure}[]{0.23\linewidth}
  \centering
  % include second image
  \includegraphics[width=.6\linewidth]{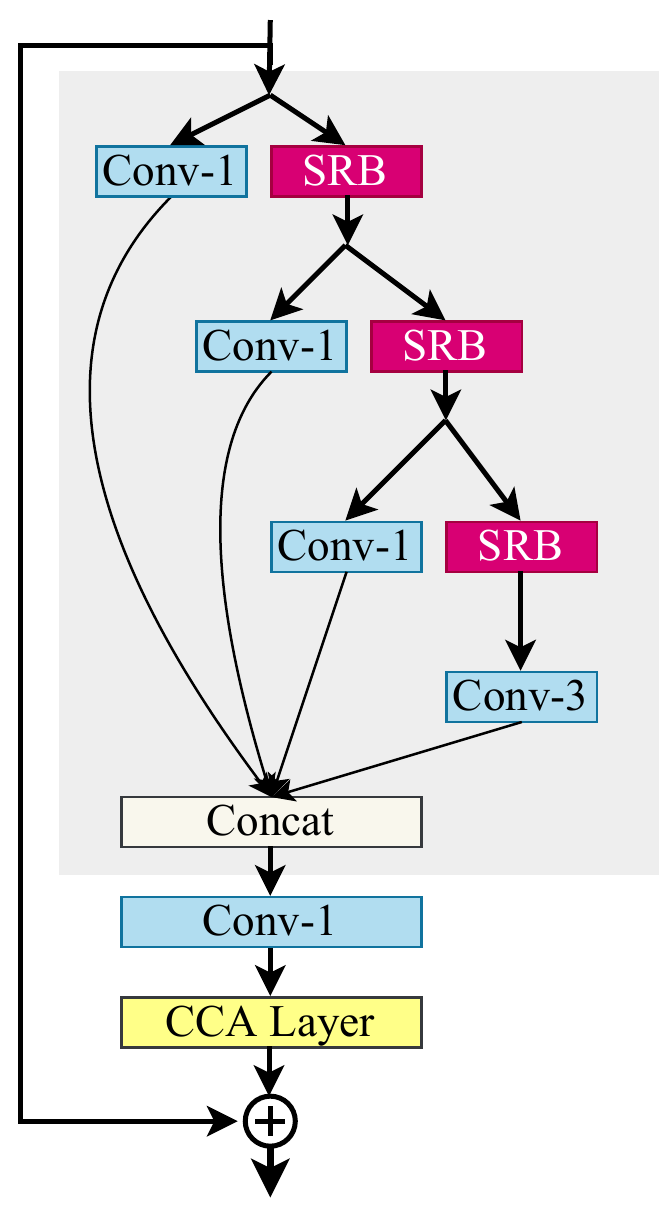}  
  \caption{RFDB}
  \label{fig:imdc_srb}
\end{subfigure}
\caption{The Base block, SRB block, FDC block and RFDB used in ablation study.}
\label{fig:ablation_net}
\end{figure*}

\section{EXPERIMENTS}
\subsection{Datasets and metrics}
Following previous works~\cite{EDSR,RCAN,IDN,CARN,IMDN}, we use the recently popular dataset DIV2K~\cite{NTIRE2017} to train our models. The DIV2K dataset contains
800 high-quality RGB training images. For testing, we use five widely used benchmark datasets: Set5~\cite{Set5}, Set14~\cite{Set14},
BSD100~\cite{B100}, Urban100~\cite{Urban100} and Manga109~\cite{Manga109}. We employ peak signal-to-noise ratio (PSNR) and structural similarity (SSIM)~\cite{SSIM}
to measure the quality of the super-resolved images. All the values are calculated on the Y channel of the 
YCbCr channels converted from the RGB channels as with existing works~\cite{VDSR,DRRN,EDSR,RDN,RCAN,CARN,IDN,IMDN}.

\subsection{Implementation details}\label{sec:imp}
We generate the training LR images by down-sampling HR images with scaling factors 
($\times2$, $\times 3$ and $\times 4$) using bicubic interpolation in MATLAB. As of preparing for this paper,
the IMDN has not released the training code yet. To reproduce the results that reported in the IMDN paper, we
use different training settings from the original paper. More details will be discussed in section~\ref{sec:exp}.
In this paper, we randomly crop 64 patches of size $64\times 64$ from the LR images as input for each training
minibatch. We augment the training data with random horizontal flips and 90 rotations. We train our model with 
ADAM optimizer by setting $\beta_1 = 0.9$, $\beta_2=0.999$, and $\epsilon=10^{-8}$. The learning rate is
initialized as $5\times 10^{-4}$ and halved at every $2\times 10^{5}$ minibatch updates. When training the final
models, the $\times 2$ model is trained from scratch. After the model converges, we use it as a pretrained network
for other scales. All the models in the ablation study are trained from scratch for
saving the training time. 
We implement two models in this paper, which are named RFDN and RFDN-L. RFDN uses a channel number of 48 while
RFDN-L uses a channel number of 52 to ahcieve a better reconstruction quality.
We set the number of RFDB to 6 in both RFDN and RFDN-L. 
The networks are implemented by using PyTorch framework with a NVIDIA 1080Ti GPU.

\subsection{Model analysis}
\subsubsection{Ablation study}
To evaluate the importance of the proposed feature distillation connection (FDC) and shallow residual block (SRB),
we design four blocks that will be stacked as the body part of the SR network (Figure~\ref{fig:rmdn}), respectively.
The four blocks are depicted in Figure~\ref{fig:ablation_net} and the evaluation results are shown in 
Table~\ref{tab:ablation_net}. Comparing the first two rows of Table~\ref{tab:ablation_net}, we can find that SRB
improves the performance (\eg PSNR:\textbf{+0.12dB}, SSIM:\textbf{+0.0024} for Manga109) without introducing any extra parameters.
We can also observe similar improvements when comparing the last two rows, which indicates the effectiveness of
the shallow residual block. By adding FDC, the performance of the base method is improved by a large margin,
for example the PSNR of Manga109 improves from 30.28 to 30.47 (\textbf{+0.19dB}). Thanks to FDC and SRB, our RFDB
significantly outperforms the base block.
\begin{table}[t]
    \centering
    \caption{Investigations of FDC and SRB on the benchmark datasets with scale factor of $\times 4$.
    The best results are highlighted.}
    \resizebox{0.8\linewidth}{!}{
    \begin{tabular}{lcccccc}
        \toprule
        Method & Params & Set5 & Set14 & B100 & Urban100 & Manga109\\
        \midrule
        Base & 652K & 32.08/0.8932 & 28.55/0.7802 & 27.53/0.7345 & 26.05/0.7842 & 30.28/0.9050\\ 
        SRB & 652K & \textbf{32.19}/0.8949 & 28.58/0.7809 & 27.53/0.7347 & 26.07/0.7849 & 30.40/0.9074\\
        FDC & 637K & 32.18/0.8945 & 28.58/0.7811 & 27.55/0.7352 & 26.09/0.7849 & 30.47/0.9077\\
        RFDB & 637K & 32.18/\textbf{0.8950} & \textbf{28.61}/\textbf{0.7820} & \textbf{27.56}/\textbf{0.7356} & \textbf{26.10}/\textbf{0.7859} & \textbf{30.55}/\textbf{0.9082}\\
        \bottomrule
    \end{tabular}}
    \label{tab:ablation_net}
\end{table}
\begin{table}[t]
    \centering
    \caption{Investigations of the distillation rate on the benchmark datasets with scale factor of $\times 4$.
    The best results are highlighted. $\uparrow$ represents rising, $\downarrow$ represents falling and $\wedge$
    represents rising first and then falling.}
    \resizebox{0.8\linewidth}{!}{
    \begin{tabular}{lcccccc}
        \toprule
        Ratio & Params & Set5 & Set14 & B100 & Urban100 & Manga109\\
        \midrule
        0.25 & 523K & \textbf{32.18}/\textbf{0.8946} & 28.57/0.7811 & 27.53/0.7348 & 26.09/0.7851 & 30.44/0.9071\\
        0.5 & 544K & 32.16/0.8945 & 28.60/\textbf{0.7819} & \textbf{27.55}/\textbf{0.7351} & 26.10/\textbf{0.7858} & 30.45/0.9074\\
        0.75 & 565K & 32.15/0.8944 & \textbf{28.61}/0.7816 & 27.54/0.7350 & \textbf{26.12}/0.7853 & \textbf{30.46}/\textbf{0.9081}\\
        - &- & $\downarrow$/$\downarrow$ & $\uparrow$/$\wedge$ & $\wedge$/$\wedge$ & $\uparrow$/$\wedge$ & $\uparrow$/$\uparrow$\\ 
        \bottomrule
    \end{tabular}}
    \label{tab:mdc_rate}
\end{table}

\subsubsection{Investigation of distillation rate}
We investigate the distillation rate of the feature distillation connections in Table~\ref{tab:mdc_rate}. Different distillation rates indicate different number of output
channels in the feature distillation connections. As shown in the last row of Table~\ref{tab:mdc_rate}, when the distillation rate increases, the
growth trends of PSNR and SSIM are different on each dataset. Overall, the distillation rate of 0.5 has a good trade-off between SR performance 
and the number of parameters, which is adopted as the final distillation rate in our RFDN and RFDN-L.
\begin{table}[t]
    \centering
	\begin{minipage}{0.45\linewidth}
		\centering
        \includegraphics[width=0.8\linewidth]{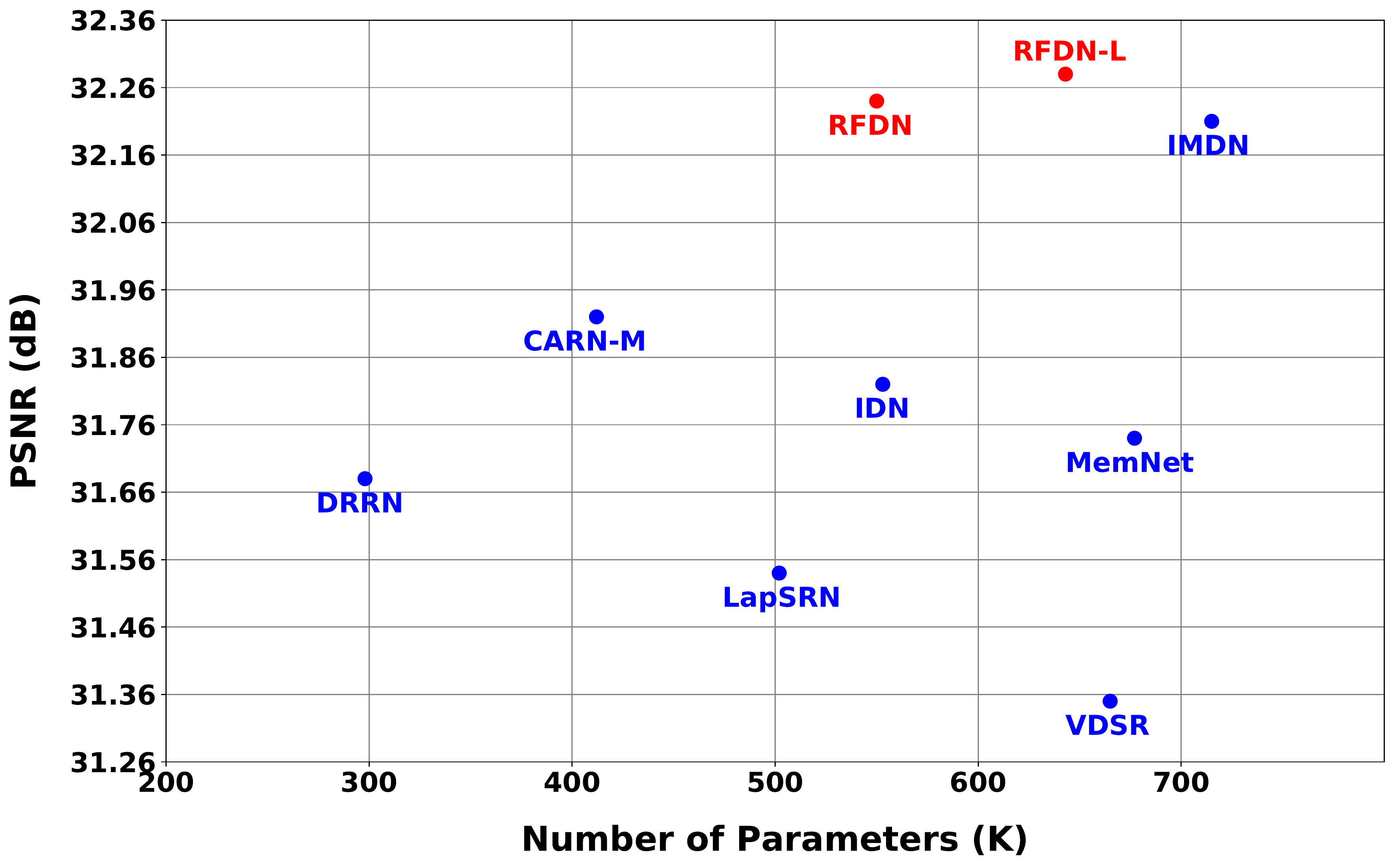}  
		\captionof{figure}{PSNR \vs Parameters.}
		\label{fig:param}
	\end{minipage}
	\begin{minipage}{0.45\linewidth}
		\centering
        \includegraphics[width=0.8\linewidth]{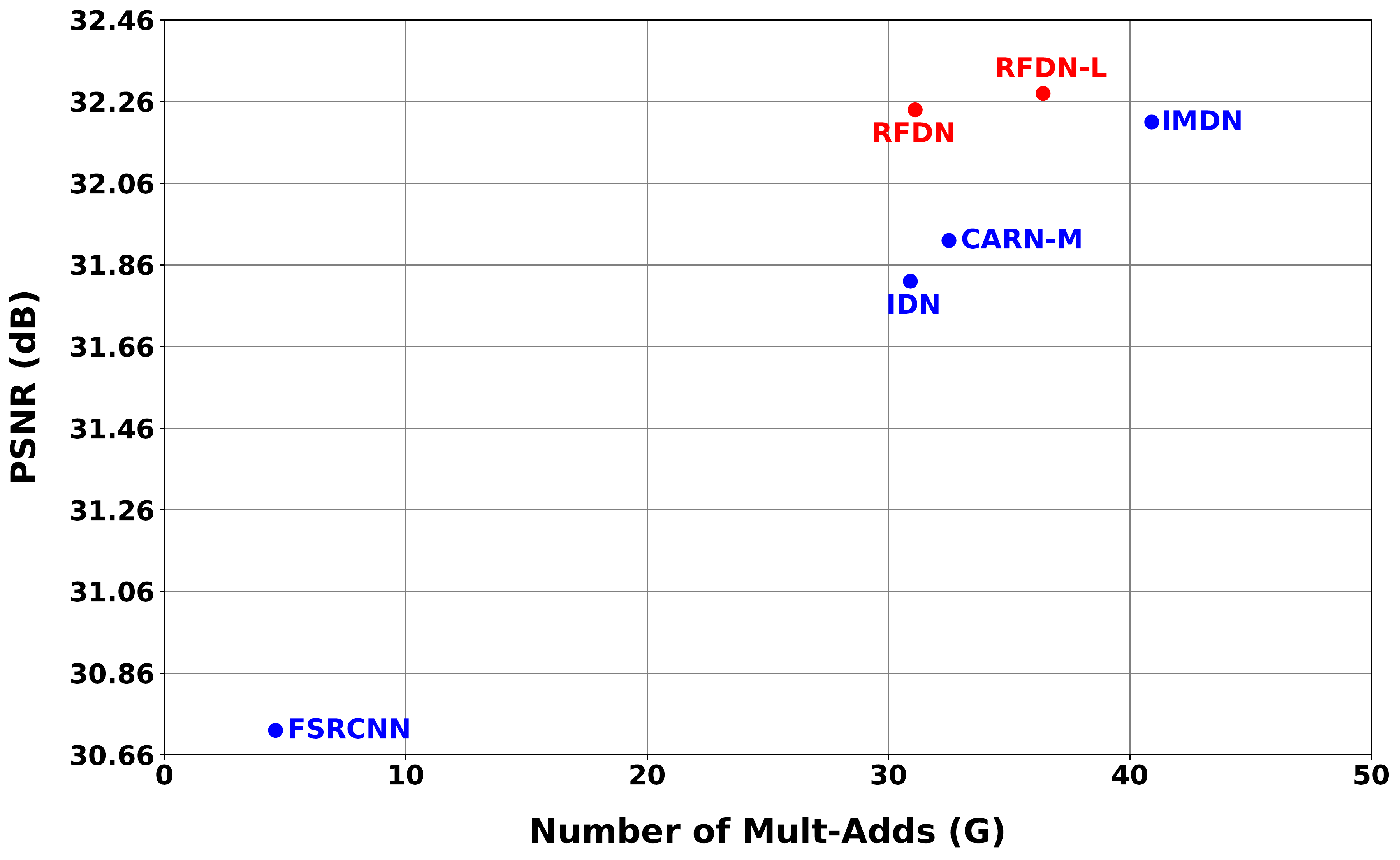}  
		\captionof{figure}{PSNR \vs Mult-Adds.}
		\label{fig:mult_add}
	\end{minipage}
\end{table}

\subsubsection{Model complexity analysis}
Figure~\ref{fig:param} depicts the comparison of PSNR \vs parameters on Set5 $\times 4$ dataset. 
The models depicted in Figure~\ref{fig:param} including DRRN~\cite{DRRN}, LapSRN~\cite{LapSRN}, VDSR~\cite{VDSR}, MemNet~\cite{MemNet}, IDN~\cite{IDN}, CARN-M~\cite{CARN} and IMDN~\cite{IMDN}. 
When evaluating a lightweight model, 
the number of model parameters is a key factor to take into account. From Table~\ref{tab:sota}, we can observe
that our RFDN achieves comparable or better performance when comparing with the state-of-the-art lightweight
models with fewer parameters. As shown in Figure~\ref{fig:param},
though IMDN achieves prominent improvements compared with the previous
methods, such as MemNet and IDN, it has more parameters than most of the lightweight models. In contrast, our
RFDN achieves better performance than VDSR, MemNet, IDN, and IMDN with fewer parameters. 
When using more feature channels, our RFDN-L achieves even better results than RFDN while 
maintaining a modest model size. To get a more comprehensive understanding of the model complexity, we also show
the comparison of PSNR \vs Mult-Adds on Set5 $\times 4$ dataset in Figure~\ref{fig:mult_add}. As we can see,
our RFDN and RFDN-L achieve higher PSNR than IMDN while using fewer calculations. IMDN won the first place in 
the parameters and inference tracks of AIM 2019 constrained super-resolution challenge~\cite{AIM2019}, so we compare our RFDN 
with IMDN in terms of FPS. Our RFDN (44 FPS) has a comparable inference speed with IMDN (49 FPS) while being
more accurate and lightweight. Moreover, our method has fewer calculations than IMDN and can save more energy.
%############################################################################################################
%\begin{figure*}[t]
%    \centering
%  \includegraphics[width=0.4\linewidth]{Figure/set5_x4_param.pdf}  
%  \includegraphics[width=0.4\linewidth]{Figure/set5_x4_multadds.pdf}  
%  \caption{Trade-off  between performance and number of parameters on Manga109 $\times 4$ dataset.}
%  \label{fig:param}
%\end{figure*}
%############################################################################################################
\begin{table}[t]
    \centering
    \caption{Average PSNR/SSIM for scale factor 2, 3 and 4 on datasets Set5, Set14, BSD100, Urban100, and Manga109. The best and second best results are highlighted in {\color{red}red} and {\color{blue}blue} respectively.}
    \resizebox{0.6\linewidth}{!}{
    \begin{tabular}{|l|c|c|c|c|c|c|c|}
    \hline
    \multirow{2}{*}{Method} & \multicolumn{1}{l|}{\multirow{2}{*}{Scale}} & \multirow{2}{*}{Params} & Set5 & Set14 & BSD100 & Urban100 & Manga109 \\
    \cline{4-8} 
    & \multicolumn{1}{l|}{} &  & PSNR/SSIM & PSNR/SSIM & PSNR/SSIM & PSNR/SSIM & PSNR/SSIM \\
    \hline
    \hline
    Bicubic & \multirow{13}{*}{x2} & - & 33.66/0.9299 & 30.24/0.8688 & 29.56/0.8431 & 26.88/0.8403 & 30.80/0.9339 \\
    SRCNN~\cite{SRCNN} &  & 8K & 36.66/0.9542 & 32.45/0.9067 & 31.36/0.8879 & 29.50/0.8946 & 35.60/0.9663 \\
    FSRCNN~\cite{FSRCNN} &  & 13K & 37.00/0.9558 & 32.63/0.9088 & 31.53/0.8920 & 29.88/0.9020 & 36.67/0.9710 \\
    VDSR~\cite{VDSR} &  & 666K & 37.53/0.9587 & 33.03/0.9124 & 31.90/0.8960 & 30.76/0.9140 & 37.22/0.9750 \\
    DRCN~\cite{DRCN} &  & 1774K & 37.63/0.9588 & 33.04/0.9118 & 31.85/0.8942 & 30.75/0.9133 & 37.55/0.9732 \\
    LapSRN~\cite{LapSRN} &  & 251K & 37.52/0.9591 & 32.99/0.9124 & 31.80/0.8952 & 30.41/0.9103 & 37.27/0.9740 \\
    DRRN~\cite{DRRN} &  & 298K & 37.74/0.9591 & 33.23/0.9136 & 32.05/0.8973 & 31.23/0.9188 & 37.88/0.9749 \\
    MemNet~\cite{MemNet} &  & 678K & 37.78/0.9597 & 33.28/0.9142 & 32.08/0.8978 & 31.31/0.9195 & 37.72/0.9740 \\
    IDN~\cite{IDN} &  & 553K & 37.83/0.9600 & 33.30/0.9148 & 32.08/0.8985 & 31.27/0.9196 & 38.01/0.9749 \\
    %EDSR &  & 1370K & 37.99/0.9604 & 33.57/0.9175 & 32.16/0.8994 & 31.98/0.9272 & 38.54/0.9769 \\
    SRMDNF~\cite{SRMDNF} &  & 1511K & 37.79/0.9601 & 33.32/0.9159 & 32.05/0.8985 & 31.33/0.9204 & 38.07/0.9761 \\
    CARN~\cite{CARN} &  & 1592K & 37.76/0.9590 & 33.52/0.9166 & 32.09/0.8978 & 31.92/0.9256 & 38.36/0.9765 \\
    IMDN~\cite{IMDN} &  & 694K & {38.00}/{\color{blue}0.9605} & {33.63}/0.9177 & {\color{red}32.19}/{\color{red}0.8996} & {\color{blue}32.17}/{\color{blue}0.9283} & {\color{blue}38.88}/{\color{red}0.9774} \\
    \textbf{RFDN (Ours)} & & 534K & {\color{blue}38.05}/{\color{red}0.9606} & {\color{red}33.68}/{\color{blue}0.9184} & 32.16/{\color{blue}0.8994} & {32.12}/{0.9278} & {\color{blue}38.88}/{\color{blue}0.9773} \\
    \textbf{RFDN-L (Ours)} & & 626K & {\color{red}38.08}/{\color{red}0.9606} & {\color{blue}33.67}/{\color{red}0.9190} & {\color{blue}32.18}/{\color{red}0.8996} & {\color{red}32.24}/{\color{red}0.9290} & {\color{red}38.95}/{\color{blue}0.9773} \\
    \hline
    \hline
    Bicubic & \multirow{13}{*}{x3} & - & 30.39/0.8682 & 27.55/0.7742 & 27.21/0.7385 & 24.46/0.7349 & 26.95/0.8556 \\
    SRCNN~\cite{SRCNN} &  & 8K & 32.75/0.9090 & 29.30/0.8215 & 28.41/0.7863 & 26.24/0.7989 & 30.48/0.9117 \\
    FSRCNN~\cite{FSRCNN} &  & 13K & 33.18/0.9140 & 29.37/0.8240 & 28.53/0.7910 & 26.43/0.8080 & 31.10/0.9210 \\
    VDSR~\cite{VDSR} &  & 666K & 33.66/0.9213 & 29.77/0.8314 & 28.82/0.7976 & 27.14/0.8279 & 32.01/0.9340 \\
    DRCN~\cite{DRCN} &  & 1774K & 33.82/0.9226 & 29.76/0.8311 & 28.80/0.7963 & 27.15/0.8276 & 32.24/0.9343 \\
    LapSRN~\cite{LapSRN} &  & 502K & 33.81/0.9220 & 29.79/0.8325 & 28.82/0.7980 & 27.07/0.8275 & 32.21/0.9350 \\
    DRRN~\cite{DRRN} &  & 298K & 34.03/0.9244 & 29.96/0.8349 & 28.95/0.8004 & 27.53/0.8378 & 32.71/0.9379 \\
    MemNet~\cite{MemNet} &  & 678K & 34.09/0.9248 & 30.00/0.8350 & 28.96/0.8001 & 27.56/0.8376 & 32.51/0.9369 \\
    IDN~\cite{IDN} &  & 553K & 34.11/0.9253 & 29.99/0.8354 & 28.95/0.8013 & 27.42/0.8359 & 32.71/0.9381 \\
    %EDSR &  & 1555K & 34.37/0.9270 & 30.28/0.8417 & 29.09/0.8052 & 28.15/0.8527 & 33.45/0.9439 \\
    SRMDNF~\cite{SRMDNF} &  & 1528K & 34.12/0.9254 & 30.04/0.8382 & 28.97/0.8025 & 27.57/0.8398 & 33.00/0.9403 \\
    CARN~\cite{CARN} &  & 1592K & 34.29/0.9255 & 30.29/0.8407 & 29.06/0.8034 & 28.06/0.8493 & 33.50/0.9440 \\
    IMDN~\cite{IMDN} &  & 703K & 34.36/0.9270 & 30.32/0.8417 & {\color{blue}29.09}/0.8046 & 28.17/0.8519 & 33.61/0.9445 \\
    \textbf{RFDN (Ours)} & & 541K & {\color{blue}34.41}/{\color{blue}0.9273} & {\color{blue}30.34}/{\color{blue}0.8420} & {\color{blue}29.09}/{\color{red}0.8050} & {\color{blue}28.21}/{\color{blue}0.8525} & {\color{blue}33.67}/{\color{blue}0.9449} \\
    \textbf{RFDN-L (Ours)} & & 633K & {\color{red}34.47}/{\color{red}0.9280} & {\color{red}30.35}/{\color{red}0.8421} & {\color{red}29.11}/{\color{red}0.8053} & {\color{red}28.32}/{\color{red}0.8547} & {\color{red}33.78}/{\color{red}0.9458} \\
    \hline
    \hline
    Bicubic & \multirow{13}{*}{x4} & - & 28.42/0.8104 & 26.00/0.7027 & 25.96/0.6675 & 23.14/0.6577 & 24.89/0.7866 \\
    SRCNN~\cite{SRCNN} &  & 8K & 30.48/0.8626 & 27.50/0.7513 & 26.90/0.7101 & 24.52/0.7221 & 27.58/0.8555 \\
    FSRCNN~\cite{FSRCNN} &  & 13K & 30.72/0.8660 & 27.61/0.7550 & 26.98/0.7150 & 24.62/0.7280 & 27.90/0.8610 \\
    VDSR~\cite{VDSR} &  & 666K & 31.35/0.8838 & 28.01/0.7674 & 27.29/0.7251 & 25.18/0.7524 & 28.83/0.8870 \\
    DRCN~\cite{DRCN} &  & 1774K & 31.53/0.8854 & 28.02/0.7670 & 27.23/0.7233 & 25.14/0.7510 & 28.93/0.8854 \\
    LapSRN~\cite{LapSRN} &  & 502K & 31.54/0.8852 & 28.09/0.7700 & 27.32/0.7275 & 25.21/0.7562 & 29.09/0.8900 \\
    DRRN~\cite{DRRN} &  & 298K & 31.68/0.8888 & 28.21/0.7720 & 27.38/0.7284 & 25.44/0.7638 & 29.45/0.8946 \\
    MemNet~\cite{MemNet} &  & 678K & 31.74/0.8893 & 28.26/0.7723 & 27.40/0.7281 & 25.50/0.7630 & 29.42/0.8942 \\
    IDN~\cite{IDN} &  & 553K & 31.82/0.8903 & 28.25/0.7730 & 27.41/0.7297 & 25.41/0.7632 & 29.41/0.8942 \\
    %EDSR &  & 1518K & 32.09/0.8938 & 28.58/0.7813 & 27.57/0.7357 & 26.04/0.7849 & 30.35/0.9067 \\
    SRMDNF~\cite{SRMDNF} &  & 1552K & 31.96/0.8925 & 28.35/0.7787 & 27.49/0.7337 & 25.68/0.7731 & 30.09/0.9024 \\
    CARN~\cite{CARN} &  & 1592K & 32.13/0.8937 & {\color{blue}28.60}/0.7806 & {\color{red}27.58}/0.7349 & 26.07/0.7837 & 30.47/{0.9084} \\
    IMDN~\cite{IMDN} &  & 715K & {32.21}/{0.8948} & 28.58/0.7811 & 27.56/0.7353 & 26.04/0.7838 & 30.45/0.9075 \\
    \textbf{RFDN (Ours)} & & 550K & {\color{blue}32.24}/{\color{blue}0.8952} & {\color{red}28.61}/{\color{red}0.7819} & {\color{blue}27.57}/{\color{blue}0.7360} & {\color{blue}26.11}/{\color{blue}0.7858} & {\color{blue}30.58}/{\color{blue}0.9089} \\
    \textbf{RFDN-L (Ours)} & & 643K & {\color{red}32.28}/{\color{red}0.8957} & {\color{red}28.61}/{\color{blue}0.7818} & {\color{red}27.58}/{\color{red}0.7363} & {\color{red}26.20}/{\color{red}0.7883} & {\color{red}30.61}/{\color{red}0.9096} \\
    \hline

    \end{tabular}}
    \label{tab:sota}
\end{table}
\begin{table*}[t]
    \centering
    \caption{Performance comparison of RFDN and IMDN under the same experimental settings. Both models are trained
    from scratch with scaling facotr $\times 4$.}
    \resizebox{0.8\linewidth}{!}{
    \begin{tabular}{lcccccc}
        \toprule
        Method & Params & Set5 & Set14 & B100 & Urban100 & Manga109\\
        \midrule
        IMDN~\cite{IMDN} & 715K & 32.16/0.8940 & {28.59}/0.7812 & {27.54}/0.7350 & 26.05/0.7841 & 30.42/0.9074\\ 
        RFDN & 550K & \textbf{32.24}/\textbf{0.8953} & \textbf{28.59}/\textbf{0.7814} & \textbf{27.54}/\textbf{0.7355} & \textbf{26.15}/\textbf{0.7868} & \textbf{30.48}/\textbf{0.9080}\\
        \bottomrule
\end{tabular}}
\label{tab:exp_setup}
\end{table*}
%\begin{figure}
%    \centering
%  \includegraphics[width=0.75\linewidth]{Figure/demo_2.pdf}  
%  \caption{Visual comparisons of RFDN with other SR methods on Urban100 $\times4$ dataset.}
%  \label{fig:demo}
%\end{figure}

\subsection{Comparison with state-of-the-arts}
We compare the proposed RFDN with various lightweight SR methods on $\times 2$, $\times3$ and $\times 4$ scales,
including SRCNN~\cite{SRCNN}, FSRCNN~\cite{FSRCNN}, VDSR~\cite{VDSR}, DRCN~\cite{DRCN}, LapSRN~\cite{LapSRN}, DRRN~\cite{DRRN}, MemNet~\cite{MemNet}, IDN~\cite{IDN}, SRMDNF~\cite{SRMDNF}, CARN~\cite{CARN} and IMDN~\cite{IMDN}. Table~\ref{tab:sota} shows the quantitative comparisons
on the five benchmark datasets. We can find that the proposed RFDN can make a better trade-off than IMDN. Our RFDN can achieve comparable or better
results with state-of-the-art methods while using 534/541/550K parameters for $\times 2$/$\times 3$/$\times 4$ SR. By using slightly
more parameters, our RFDN-L achieves the best in most quantitative results, especially on large scaling factors.
%Figure~\ref{fig:demo} shows the visual comparisons on Urban100 $\times 4$ dataset. Our RFDN can recover better
%local details than others, which also demonstrates the superiority of our method.

\subsection{About the experimental settings}\label{sec:exp}

As described in section~\ref{sec:imp}, we use a slightly different experimental setup when training our models.
In order to get a clearer insight on the improvements of our RFDN, we train both RFDN and the IMDN~\cite{IMDN} from scratch
under the same experimental settings. Table~\ref{tab:exp_setup} shows the performance comparison on the five
benchmark datasets. Our RFDN outperforms IMDN on all the datasets in terms of both PSNR and SSIM with much fewer
parameters, which proves that the improvements on network design of our RFDN indeed boosts
the performance of image SR.

\subsection{Enhanced RFDN for AIM20 challenge}
As shown in Table~\ref{tab:AIM}, our enhanced RFDN (E-RFDN) won the first place in the AIM 2020 efficient super-resolution challenge~\cite{AIM20}. 
Specifically, we replace the CCA layer in RFDB with the ESA block~\cite{RFANet} and we use 4 such enhanced RFDBs (E-RFDBs) in E-RFDN.
The number of feature channels in E-RFDN is set to 50 and the feature distillation rate is 0.5. During the training of E-RFDN, 
HR patches of size $256\times 256$ are randomly cropped from HR images, and the mini-batch size is set to 64. The E-RFDN model is trained
by minimizing L1 loss function with Adam optimizer. The initial learning rate is set to $5\times 10^{-4}$ and halved at every
200 epochs. After 1000 epochs, L2 loss is used for fine-tuning with learning rate of $1\times10^{-5}$. 
DIV2K and Flickr2K datasets are used for training the E-RFDN model. We include the top five methods in Table~\ref{tab:AIM}, the ``\#Activations'' 
measures the number of elements of all outputs of convolutional layers. Compared to the first place method IMDN in AIM 2019 constrained SR challenge~\cite{AIM19}, our method provides a significant gain with respect to the runtime, parameters, FLOPs, and activations.
More details and reuslts can be found in~\cite{AIM20}.

\begin{table*}[!htbp]\footnotesize
    \caption{AIM 2020 efficient SR challenge results (we only include the fisrt five methods).} %\vspace{-0.25cm}
\center
    \resizebox{0.8\linewidth}{!}{
\begin{tabular}{p{3cm}|p{1.7cm}||p{0.9cm}<{\centering}|p{1.7cm}<{\centering}||p{1.7cm}<{\centering}|p{1.7cm}<{\centering}|p{1.9cm}<{\centering}|p{0.9cm}<{\centering}}
    \multirow{2}{*}{Team} &  \multirow{2}{*}{Author}  & PSNR & Runtime & \#Params. & FLOPs & \#Activations  & Extra \\%\cline{4-10}
                          &  &[test] & [s] & [M]  & [G]  & [M]& Data \\\hline\hline
    
    NJU\_MCG (ours) &	TinyJie & 28.75	& 0.037 & 0.433	&	27.10 & 112.03  &	Yes\\
AiriA\_GG &	Now & 28.70	& 0.037 & 0.687	& 44.98 & 118.49 &	Yes\\
UESTC-MediaLab & Mulns & 28.70 & 0.060 & 0.461	& 30.06 & 219.61 &	Yes\\
XPixel & zzzhy & 28.70 & 0.066 & 0.272	& 32.19 & 270.53 &	Yes\\
HaiYun & Sudo & 28.78 & 0.058 & 0.777	& 49.67 & 132.31 &	Yes\\
\hline
IMDN	&zheng222&	28.78	& 0.050	& 0.893	& 58.53	& 154.14	&Yes\\
Baselin &MSRResNet&	28.70 &	0.114 & 1.517 & 166.36 & 292.55	& Yes\\
\hline
\end{tabular}}
\label{tab:AIM}
\end{table*}

\section{CONCLUSION}
In this paper, we give a comprehensive analysis of the information distillation mechanism for lightweight image super-resolution. Then we rethink the
information multi-distillation network (IMDN) and propose the feature distillation connections (FDC) that are much more lightweight and flexible. To further boost the
super-resolution performance, we also propose the shallow residual block (SRB) that incorporates the identity connection with one convolutional block.
By using the shallow residual blocks and the feature distillation connections, we build the residual feature distillation network (RFDN) for fast and 
lightweight image super-resolution. Extensive experiments have shown that the proposed method achieves state-of-the-art results both 
quantitatively and qualitatively. Furthermore, our model has a modest number of parameters and mult-adds such that it can be easily ported 
to mobile devices.

%\clearpage\mbox{}Page \thepage\ of the manuscript.
%\clearpage\mbox{}Page \thepage\ of the manuscript.

%This is the last page of the manuscript.
%\par\vfill\par
%Now we have reached the maximum size of the ECCV 2020 submission (excluding references).
%References should start immediately after the main text, but can continue on p.15 if needed.

%\clearpage
\bibliographystyle{splncs04}
\bibliography{egbib}
\end{document}